

\magnification= \magstep1
\tolerance=1600
\parskip=5pt
\baselineskip= 6 true mm

\def\a{\alpha}
\def\b{\beta}
\def\g{\gamma} \def\G{\Gamma}

\def\l{\lambda} \def\L{\Lambda}
\def\m{\mu}
\def\n{\nu}
\def\j{\psi} 

\def\x{\xi} 
\def\w{\omega} \def\W{\Omega}
\def\k{\kappa}
\def\dd{{\rm d}}
\def\cl{\centerline}
\def\pa{\partial}
\def\R{I\!\!R}
\def\E{I\!\!I}
\def\half{{\textstyle{1\over2}}}
{\nopagenumbers

\vglue 1truecm
\rightline{THU-94/18}
\rightline{hep-th/9411228}
\rightline{December 1994}
\vfil
\cl{\bf LATTICE REGULARIZATION OF GAUGE THEORIES}
 \smallskip \cl{\bf WITHOUT LOSS OF CHIRAL SYMMETRY}
\vskip 1 truecm
\cl{G. 't Hooft}
\vskip 1 truecm
\cl{Institute for Theoretical Physics}
\cl{University of Utrecht, P.O.Box 80 006}
\cl{3508 TA Utrecht, the Netherlands}
\vfil
\noindent{\bf Abstract}\smallskip

A lattice regularization procedure for gauge theories is proposed in
which fermions are given a special treatment such that all chiral
flavor symmetries that are free of Adler-Bell-Jackiw anomalies are kept
intact. There is no doubling of fermionic degrees of freedom. A price
paid for this feature is that the number of fermionic degrees of
freedom per unit cell is still infinite, although finiteness of the
complete functional integrals can be proven (details are outlined in an
Appendix).  Therefore, although perhaps of limited usefulness for
numerical simulations, our scheme can be applied for studying aspects
such as analytic convergence questions, spontaneous symmetry breakdown
and baryon number violation in non-Abelian gauge theories.

\vfil\eject}
\noindent{\bf 1. Introduction.}\smallskip

Conservation of chiral symmetries in lattice gauge theories is often
considered to be an important problem. One would like to study for
instance the spontaneous breakdown of chiral $SU(2)\times SU(2)$ in QCD
but this study is complicated by the fact that in the Wilson action [1]
such symmetries are also explicitly broken [2]. Remedies of this
problem were often sought in alternative versions of this action[3],
but then it was found that either other subgroups of $SU(2)\times
SU(2)$ were sacrificed or else extra fermionic degrees of freedom
emerged as unwanted modes of the fermionic action.

The origin of this difficulty was not difficult to identify. Chiral
symmetries are offset by anomalies so that the actual symmetries of the
system are smaller than the apparent symmetries of the Lagrangian.
Hence, the fact that the symmetries have to be smaller than that of the
unregularized Lagrangian should not be a surprise. The problem that
remains however is that the chiral symmetries are broken more than
necessary. The anomaly only removes the chiral $U(1)$ part of this
symmetry whereas most lattice actions remove the chiral part of $U(2)
\times U(2)$ altogether.

No-go theorems were then formulated [4] stipulating that this is a
nuisance that will stay with us forever. But these theorems used
certain assumtions concerning the fermionic part of the action, and, as
is often the case with such theorems, its conclusion is no worse than
the assumption. All that has to be done is be more imaginative in the
fermion sector.

In this paper, our essential trick consists of observing that the
fermions need to know how the instanton winding numbers are defined on
the lattice plaquettes and hypercubes. If one would add a tiny
instanton somewhere within one hypercube this would add a fermionic
zero mode, or equivalently, an other chiral fermionic degree of freedom
inside that hypercube. We will fix these winding numbers simply by
extending the gauge fields, originally only defined on the lattice
links, in the smoothest possible way to all space-time points within as
well as on the edges of each hypercube. This procedure has no effect on
the gauge degrees of freedom, which will always be represented just by
the connections on the links, but it will affect the interactions with
the fermions.

The fermionic part of the action is now defined by having the fermions live
in this continuous space-time. We then regularize the fermions separately,
using for instance a Pauli-Villars procedure (although any other
regularization procedure with the appropriate adjustments of the
finite subtraction terms will do). What makes this procedure attractive is
that the fermionic functional integral is merely a functional determinant
which in principle is easy to compute. The prescription is thus to
regularize this determinant first. An apparent difficulty is that the
fermions live in a ``background gauge field'' (which is generated by the
gauge link variables $U_\m(x)$ and is to be integrated over later), and
this field may be fairly strong. We now {\it prove} that in spite of this
the regularization procedure is completely convergent for all $U_\m$
configurations. This proof is outlined in the Appendix.

\bigbreak
\noindent{\bf 2. The gauge field in between lattice sites.}\smallskip

The gauge field part of our lattice theory is as usual. On a cubic
lattice the gauge field degrees of freedom are chosen to be elements
$U_\m(x)$ of the gauge group $G$, situated on the unit links connecting
the space-time points $x$ and $x+e_\m$. The gauge field action is also
chosen to be the usual thing:  $$S_{YM}={1\over g^2}\sum_{x,\m,\n}{\rm
tr}\Big(U_\m(x)U_\n(x+e_\m)U^\dagger_\m(x+e_\n)
U^\dagger_\n(x)\Big)\,.\eqno(1)$$

Now before being able to introduce the fermion fields we must extend
our definition of the gauge field to all other space-time points. Thus,
although in Eq. (1) the points $x$ are integers denoting the lattice
sites, we now choose points $x\in \R^4$ to define a vector connection
field $A_\m(x)$ in all these points.

First we define $A_\m(x)$ for the points $x$ on the lattice links
$(x,\m)$, simply by $$U_\m(x)\equiv e^{iaA_\m(x)}\,,\eqno(2)$$ where
$a$ is the lattice length, and $A_\m$ is given as an element of the Lie
algebra of $G$. It will turn out to be important to observe that Eq.
(2) only defines $A_\m(x)$ uniquely if we insist that the eigenvalues
of $aA_\m(x)$ lie in the interval $(-\pi,\pi]$. We will call these
elements of the Lie algebra {\it minimal}.

Next, the gauge field on the elementary plaquettes (2-simplexes) can
also be defined in a straightforward manner. For a given plaquette
$(x,\m,\n)$ first pick the gauge in which
$$U_\m(x)=U_\n(x)=U_\m(x+e_\n)=\E\,,\quad U_\n(x+e_\m)=e^{iaA^1}\,,
\eqno(3)$$
where again $aA^1$ is chosen to be minimal. Then in that gauge
$$A_\n(x+\a e_\m +\b e_\n)\equiv {\a\over a}A^1\,,\quad A_\m(x+\a
e_\m +\b e_\n)\equiv 0\,,\quad \a,\,\b\in[0,a]\,.\eqno(4)$$
With this choice the gauge field connection on the plaquette is constant:
$$F_{\m\n}(x+\a e_\m +\b e_\n)={A^1\over a}\,,\eqno(5)$$
and it is not hard to verify that gauge transformations $\W$ of the form
$$\W(x+\a e_\m +\b e_\n)=e^{if(\a,\b)A^1}\,\eqno(6)$$
enable us to permute the four links of the plaquette in the gauge
condition (3). Thus the prescription (4) (including the minimality
condition) is a symmetric one.

Another way to see that this prescription is symmetric is by observing
that it obeys the field equation in 2-space:
$$\sum_{i\rm\ in\ plaquette\ 2-space}D_iF_{ij}(x)\,=\,0\,,\eqno(7)$$
where $D_i$ stands for the covariant derivative,
and it is the unique solution to the requirement that the 2-dimensional
action on the plaquette, $\int\dd^2x \sum_{i,j\,\rm\in\ plaquette\,}
F_{ij}F_{ij}$ has the smallest possible value, given the boundary
conditions (3) on the surrounding links.

Extending the gauge field inside a cube (3-simplex) $T^3$ is slightly
less straightforward. There are in principle many possibilities, and
the exact choice made is of lesser importance apart from one crucial
condition: we must again impose a minimality condition on the one hand,
and on the other we insist that the fields be connected continuously. A
good example is the following prescription:

\smallskip{\narrower{\noindent Choose the fields in a three-simplex to
obey the sourceless field equations of that three space:
$$\sum_{a\rm\ in\ space\ of\ 3-simplex }D_aF_{ab}\,=\,0\,,\eqno(8)$$
and of all possible solutions choose the one that minimises the
Euclidean value of $\int\dd^3x\sum_{a,b\rm\ in\ space\ of\ 3-simplex}
F_{ab}F_{ab}$, under the boundary condition (4) on the surrounding
plaquettes.}\smallskip}

\noindent This is the most natural generalisation for the 3-simplex of
our prescription on the 2-simplex, and now also the field continuation
onto the 4-simplexes $T^4$ is evident:  {\smallskip\narrower {\noindent
Choose the fields in a four-simplex to obey the usual sourceless field
equations $$\sum_{\m=1}^4D_\m F_{\m\n}\,=\,0\,,\eqno(9)$$ and again of
all possible solutions the one that minimises the Euclidean value of
$\int\dd^4x\sum_{\m\n}F_{\m\n}F_{\m\n}$, with the fields constructed
earlier on the surrounding 3-simplexes as boundary conditions.}
\smallskip}

\noindent Indeed the conditions (7), (8) and (9) ensure that our gauge
fields are extended in the smoothest possible way to all space-time
points.  Of course the derivatives of (some of) the field components
are discontinuous on the lattice links, plaquettes and 3-simplexes, but
not more than needed to obey the other requirement, {\it i.e.} that the
lattice link variables $U_\m(x)$ as independent integration variables
in the Euclidean lattice path integral coincide with
$\exp(\int_{\x\in{\rm\, link\,}(x,\m)}iA_\m(\x)\dd \x^\m)$. Therefore
(7), (8) and (9) do not hold on the lower-dimensional edges where they
are corrected by Dirac delta distributions.

Note that the exercise carried out in this section has no effect on the
pure gauge field action which is still given by Eq. (1), or on the pure
gauge field functional integral. The only reason for this exercise is
that it will be needed for the fermionic part of the gauge theory.
\bigbreak
\noindent{\bf 3. Fermions.}\smallskip

The fermions in our theory are sensitive to the gauge field topology.
This is why the Wilson action for the fermions gives problems. The
conventional lattice formulation leaves holes between the lattice sites
and so the topological winding numbers are ill-defined. In our present
formalism this problem is cured. We fixed the topology by specifying
what our gauge fields are between the lattice points and links. Our
minimality condition in eqs. (7), (8) and (9) implies that usually the
topological winding numbers there are kept at a minimum. This does not imply
that these winding numbers vanish. If we integrate over several lattice
sites we can easily find appreciable cumulation of topological winding
numbers, depending on the values of the gauge field integrands $U_\m(x)$.
There will be monopoles and instantons, but their sizes must be larger
than the lattice lengths $a$.

We have not found a discrete lattice version of the fermionic part of the
action, such that fermionic degrees of freedom are as discrete as the gauge
field variables. The point however that we wish to stress in this paper is
that such a descrete fermionic action is not at all necessary for our
theory to be properly and completely regularized. We simply propose to
keep as our fermionic action
$$S_{\rm fermions}=-\int\dd^4x\,\bar\j(x)\big(\g_\m (\pa_\m+iA_\m(x)
+m\big)\j(x)\,,\eqno(10)$$
where now $A_\m(x)$ stands for the extended gauge field constructed in the
previous section.

Of course this fermionic action still suffers from an ultraviolet
infinity. However, it is merely a one-loop infinity that is quite
easy to regularize separately. Just introduce the familiar Pauli-Villars
regulator fields, {\it i.e.} massive spinor fields with masses $\L_i$
and weights $e_i = \pm 1$. The $i=0$ component is the physical field
with $\L_0=m$ and $e_0=1$, where $m$ may or may not be zero, as dictated
by whatever chiral symmetry one would like to impose. Of the other fields
the ones with $e_i=-1$ have indefinite metric, but this does not matter
since we let all $\L_i\rightarrow\infty$. Throughout the limiting process
one imposes the identities
$$\sum_i e_i\ \equiv\ \sum_i e_i\L_i\ \equiv\ \sum_i e_i\L_i^2\ \dots\
\equiv0\,,\eqno(11)$$
where usually no higher powers than the fourth are needed for all diagrams
to converge. Furthermore
$$\sum_i e_i\log\L_i\,\equiv\,\log\L\,;\quad\sum_i e_i \L^n\log\L_i\,
\equiv\,0\,,\quad 0<n\le 4\,.\eqno(12)$$
If $\L$ in here would be kept finite then for all
field configurations $A_\m(x)$ the regularized fermionic determinant,
$$\prod_i\Big(\det(\g D+\L_i)\Big)^{e_i}\,,\eqno(13)$$
is known to be absolutely finite {\it in perturbation theory}. In
particular it is well known that in the limit where the $\L_i$ are sent
to infinity all chiral symmetries are restored with the exception of the
ones that suffer explicitly from an anomaly. In our appendix we show
that the same statements are true if we calculate the regularized
fermionic determinant {\it exactly} by myltiplying the eigenvalues
of the Dirac operator. The limit $\L_i\rightarrow\infty$ is finite.
\bigbreak
{\noindent\bf 4. The lattice gauge action. Conclusion.}
\smallskip

{}From the above it should now be clear how to prescribe rigorously a
gauge theory free from infinities and with all anomaly-free chiral
symmetries preserved. The combined action from Eqs (1) and (10),
$S_{YM}+ S_{\rm fermions}$ is {\it first} integrated over the
continuous fermion variables using the Pauli-Villars regulators. Then
we send $\L_i\rightarrow\infty$. This
yields a finite fermionic determinant, which however still depends on
the link variables $U_\m(x)$. The remaining integrand depends only on
these variables, which form a finite-dimensional space so that standard
approximation techniques may be applied to integrate over these. It is
important that before integrating over them the $\L_i$ of the regulator
fields must have been sent to infinity so that we are ensured of the
required chiral symmetries.

It is important to realise that the extrapolated fields $A_\m(x)$ play
an essential role in this construction. To see this, take a field
configuration where the $U$ variables are all close to the identity,
but still form an instanton configuration, spread over many lattice
sites.  Of course such a configuration is allowed, because instantons
scaled to large sizes have only relatively weak field values at any
specific lattice site. This configuration generates a fermionic zero
mode so that the determinant (13) vanishes unless an external source is
added. Now scale the instanton gently towards smaller sizes. The $U$
fields will become stronger, but by the time the instanton slips
between the confines of only one lattice hypercube the $U$ variables
will return towards small values again. Halfway this process, the
topological winding number must suddenly have disappeared. Thus,
although the $U$ fields changed continuously towards their new values,
the fermionic determinant must have made a jump. This is because the
extrapolated fields $A_\m(x)$ made a jump!  The discontinuity occurred
when two solutions of the equation (9) emerged for which the values of
the action $\int\dd^4xF_{\m\n}F_{\m\n}$ coincided.  The extra
requirement that we choose the solution for which this Euclidean action
is minimal forced us to jump from one solution to the other.

The fact that instantons and other topological features of our lattice
theory are constrained to be larger than the lattice length $a$ needs not
bother us. It is a lattice artifact that does not affect any of the
important symmetries, and compares well with the actual physical situation
where small instantons are suppressed because of asymptotic freedom.

Whether this prescription can be used in practical lattice calculations
is quite a different question. A perturbative calculation of the
logarithm of the determinant (in terms of one-loop diagrams, as one
usually does) will often not converge because of vanishing
eigenvalues. The determinant itself will converge in principle, but
maybe only at very high orders. Thus for large $U$ fields this
calculation may well become too cumbersome in practice for fast Monte-Carlo
simulations. One may ask whether the determinant can be approximated by
a discrete determinant and the Pauli-Villars regulators may perhaps be
replaced by some lattice scheme with a similar effect. We have to
remember however that the number of fermionic modes may vary
discontinuously as  a function of the $U$ fields, and this will pose
problems in practice. But the fundamental question whether gauge
theories can be made absolutely finite while keeping maximal ({\it
i.e.} the anomaly free part of the) chiral symmetry has been answered.
The approach advocated in this paper may well be useful for numerical
approaches towards understanding the baryon number violation phenomenon
within the Standard model, since here it is essential that the anomaly
must be the {\it only} cause for a minute symmetry breakdown, and the
question how effective this symmetry breaking will be at extremely
high energies has not yet been answered convincingly.

There is a relationship between the regularization proposed here and a
continuum regularization procedure proposed by Slavnov and Faddeev
[5].  These authors propose to regularize the continuum gauge theory by
the addition of extra derivatives in the Lagrangian. This has the
effect that all irreducible Feynman diagrams with more than one closed
loop are made convergent, except for the one-loop subdivergences, which
must be regularized first by some other method. One-loop divergences
are  easy to regularize, although in their case there are
also one-loop graphs with gauge photons going around the loops and then
a simple-minded Pauli-Villars trick leads to incorrect results. There
one can use for instance the 5-dimensional regularization scheme
proposed in Ref. [6].

\bigbreak
\noindent{\bf Acknowledgement.}\smallskip
The author thanks M. Karliner for urging him to publish this result, and
J. Smit for discussions on an ealier version of this manuscript.

\bigbreak\noindent{\bf APPENDIX:  Convergence of the regularization
procedure for the fermions.}\smallskip

We here address the following problem. Given a gauge field configuration
$A_\m(x)$, generated by our link variables $U_\m(x)$:
compute the regularised determinant of the Dirac operator
$$D\equiv \g_m\big(\pa_\m+iA_\m(x)+i\g_5 A_\m^5(x)\big)+m\,,\eqno(A.1)$$
where $m$ is a mass term that may or may not be present, $A_\m(x)$ is a
(Abelian or non-Abelian) vector potential field and $A_\m^5(x)$ an axial
vector potential field. The construction sketched in this paper that
resulted in the  gauge fields $A_\m$ and $A_\m^5$ can easily be seen to
always yield {\it bounded} values for these fields. It must be
noted that the bounds do not depend on the size of our lattice box, by
construction. Our problem is to outline a construction procedure for this
determinant, to prove that it is well-defined and finite, and to prove
that if the anomalies cancel in the one-loop 2-, 3- and 4-point diagrams,
the (chiral) flavor symmetries will be completely restored.

An essential complication is that $D$ is not Hermitean, nor can it
easily be made Hermitean by multiplying it with simple operators such as
$\g_5$. We do have that $\g_5D$ only gets a non-Hermitean contribution
from the axial gauge field. We will concentrate on computing the {\it
regularised} determinant of $F=iD$. As regulators one may use the
Pauli-Villars regulator of Eqs (11) and (12). Our argument will now go in
two steps: first we give a rigorous (non-perturbative) construction of
the regularised determinant at finite values of the regulator masses $\L_j$;
secondly we prove that the limit $\L_j\rightarrow\infty$ exists and is
approached uniformly for all $A_\m$ configurations within our bound.

Of course $F$ may have zero modes; if these are not disturbed by anything the
determinant simply vanishes, and that would be the end of the calculation.
In the other case, for doing the first computation, we need all eigenvalues
of $F$. Write
$$F\equiv K+A\quad,\quad K\equiv i\g_\m\pa_\m \,,\eqno(A.2)$$
where $K$ can be diagonalized exactly giving eigenvalues (in Euclidean space)
$\l_n=\pm|k_n|$, and $A$ is strictly bounded:
$$||A||\le \cal A\,.\eqno(A.3)$$
We now employ an important theorem concerning matrices:
\item{ } {\it Theorem:} Given an $N\times N$ matrix $F=K+A$ such that
$K$ is Hermintean and $A$ is bounded by $\cal A$ as in Eq. (A.3). Let $\k_i$
be the eigenvalues of $K$. Then the complete set of eigenvalues $\l_i$ of
$F$ (including possible degeneracies) obeys
$$|\l_i-\k_i|<C\,\cal A\,,\eqno(A.4)$$
where the coefficient $C$ may grow only logarithmically with the dimension $N$
of
the matrices.\smallskip
\noindent In fact, if $A$ is also Hermitean $C$ is one, and furthermore the
imaginary parts of $\l_i$ are bounded by $\cal A$ itself.

The proof of this theorem is somewhat lengthy. First we handle the case that
$A$ is Hermitean. This is easy, since we can write $F(x)=K+xA,\quad x\in
[0,1]$ and differentiate with respect to $x$. To handle the general
case we now split off the hermitean part of $A$, so that what remains to be
considered is the case that $A$ is purely anti-hermitean. So from now
on $A$ is anti-hermitean. One now chooses the orthonormal basis that brings
$F$ in semi-diagonal form:
\def\cd{\cdot}
$$F=\pmatrix{\l_1&\cd&\cd&\cd&\cd&\cd\cr
&\l_2&\cd&\cd&\cd&\cd\cr
&&\l_3&\cd&\cd&\cd\cr
&&&*&\cd&\cd\cr
&0&&&*&\cd\cr
&&&&&\l_N\cr}\ ,\eqno(A.5)$$
where $\l_i$ are now the eigenvalues of $F$. Let $P$ be the off-diagonal
part of $F$. Write
$$\eqalign{\l_i&=R_i+iS_i\,;\quad R_i,\ S_i\ {\rm real},\cr
K&={\rm diag}\{R_i\} + \half(P+P^\dagger)\,,\cr
A&={\rm diag}\{iS_i\} + \half(P-P^\dagger)\,,\cr
P_{k,\ell}&= 0 {\rm \quad if\quad } \ell\le k\,.\cr}\eqno(A.6) $$
The condition (A.3) implies that
$$|S_i|\le {\cal A}\,,\eqno(A.7)$$
so the imaginary parts of the eigenvalues $\l_i$ obey our bounds. Next we
deduce
from (A.7) that for all wave functions $|\psi\rangle$,
$$|\langle\psi(|P-P^\dagger)|\psi\rangle|\le 4\cal A\,,\eqno(A.8)$$
and what remains to be done is to derive from that a bound on
$|\langle\psi|(P+P^\dagger)|\psi\rangle|$. This one does as folows. Take a
wave function $|\psi^0\rangle$. Consider then
a class of normalized wave functions
$$ \psi(\w)_k= \psi^0_ke^{i\w k}\,,\eqno(A.9)$$
where $k$ is the index in the above basis representation. Define the function
$$f(\ell)=\sum_k\psi^{0*}_k\psi^0_{k+\ell}P_{k,k+\ell}\,,\eqno(A.10)$$
Then
$$\langle\psi(\w)|P|\psi(\w)\rangle=\sum_{\ell>0} f(\ell)e^{i\w\ell}\equiv
g(\w)\,.
\eqno(A.11)$$
{}From Eq. (A.8) we know that for all $\w$ there is a bound on the imaginary
part of
$g(\w)$:
$$|{\rm Im}\,g(\w)|\le 2{\cal A}\,,\eqno(A.12)$$
whereas from (A.6) we see that our theorem requires a bound on Re $g(\w)$. Now
since
according to (A.11) $g(\w)$ only has positive Fourier coefficients it obeys a
dispersion relation:
$${\rm Re}\,g(\w_1)=-{1\over\pi}{\cal P}\int\dd\w\,{{\rm Im}\,g(\w)\over
\w-\w_1}
\,,\eqno(A.13)$$ where $\cal P$ stands for principal value.
This integral is only logarithmically divergent, and that only if Im $g(\w)$
makes
a $\theta$ jump near $\w_1$. From this we derive the bound (A.4) with a
coefficient
$C$ that can diverge only logarithmically if the dimension $N$ of our matrices
becomes large.

Consider now the regularised determinant of (A.1), assuming it has no
vanishing eigenvalues. Its logarithm is
$$\G=\sum_n\sum_j e_j \log\big(\l_n+i\L_j\big)\,,\eqno(A.14)$$
where the first summand behaves for large $\l_n$ as
$$\sum_j e_j(\log\l_n+\sum_{r=1}^4{-1\over r}\left({-i\over\l_n}\right)^r\L_j^r
+{\cal O}(\L_j^5\l_n^{-5})\,.\eqno(A.15)$$
According to Eqs. (11) the logarithm term and the first
four terms in the sum can be taken to be zero. As for the remainder we
may put
$$|\l_n(A)-\k_n|<C{\cal A}\,,\eqno(A.16)$$ with $C={\cal O}(\log N)$,
where $\k_n$ are the Eigenvalues of $K$ in (A.2). For $N$
we take the number of points on a lattice that is sufficiently fine
mazed to enable us to reproduce the eigenfunction with eigenvalue $\l_n$
with reasonable accuracy. One may therefore safely say that $\cal C$
will be of order $4\log n$, when the $n^{\rm th}$ eigenvalue is
considered. One may conclude that the sum over eigenvalues $\l_n$ in
(A.15) converges uniformly.

Next we address the question of the limit where the $\L_j$ become large.
This is actually easy. We split the logarithm $\G$ up into the first
four expansion terms for small values of the $A_\m$ fields (the diagrams
with at most four external lines) and the remainder. The diagrams with
four lines or less have been calculated many times, and we know that
there are many models for which the anomalies in these diagrams can be
made to cancel out. It is these models that we are interested in.
The remainder is the set of ultraviolet convergent diagrams. The sum
over these diagrams in general does not converge (because they correspond
to the expansion of the logarithm of the determinant, which blows up
whenever there is a vanishing eigenvalue), but the sum over the
{\it regulator contributions} converges rapidly. For sufficiently
large $\L_j$ all regulator contributions for the diagrams with $k$
external lines are bounded by expressions of the form
$${1\over k}{\cal A}^k\L_j^{4-k}\,,\eqno(A.16)$$ as one can easily
convince oneself. Thus, if we sum over the $k$ values larger than 4,
(A.16) sums up to a finite bound, and we see that this contribution
vanishes in the limit $\L_j\rightarrow\infty$. So not only do these
diagrams converge, the regulator contributions for all {\it convergent}
diagrams converge uniformly to zero. We conclude that the regulator
limit of the determinant exists and obeys all wanted flavor symmetries.
\medskip
{\it Notes added:}\hfil\break
\noindent This appendix should not be regarded as a `first
proof' of finiteness of the regulated chiral determinant in an
arbitrary gauge field. I thank R.D. Ball for pointing out
references to earlier treatments of this problem [7].

That infinitely
many fermion fields would be needed on a lattice was observed in [8].
In our paper we have one chiral field, but living on a continuum,
which in practice amounts to the same thing. Narayanan and Neuberger
also explain in their paper the possible relevance of a good
regularization scheme for the study of baryon non-conservation in
the Standard Model [9].

\bigbreak
\noindent{\bf References.}\smallskip

\item{1.} K.G. Wilson, in {\it ``New Phenomena in Subnuclear Physics"}
(Edited by A. Zichichi), Plenum Press, New York (1977) (Erice Lectures,
1975).
\item{2.} L.H. Karsten and J. Smit, Nucl. Phys. {\bf B183} (1981) 103.
\item{3.} I. Montvay, Phys. Lett. {\bf B199} (1987) 89; Nucl. Phys.
(Proc. Suppl.){\bf B 4} (1988) 443; M.F.L. Golterman, Nucl. Phys.
(Proc. Suppl.) {\bf B20} (1991) 528; D.N. Petcher, Nucl. Phys. (Proc. Suppl.)
{\bf B30} (1993) 50.
\item{4.} H.B. Nielsen and M. Ninomiya, Nucl. Phys. {\bf B185} (1981) 20; Err:
Nucl. Phys. {\bf B195} (1982) 541; Nucl. Phys. {\bf B193} (1981) 173.
\item{5.} A.A. Slavnov, Theor. Math. Phys. {\bf 13} (1972) 174; {\bf
33} (1977) 210; L. Faddeev and A. Slavnov, {\it Gauge fields,
Introduction to quantum theory}, 2nd edition (Benjamin, New York, 1989).
\item{6.} G. 't Hooft, Nucl. Phys. {\bf B33} (1971) 173.
\item{7.} R.D.Ball and H. Osborn, Phys. Lett. {\bf B165} (1985) 410;
Nucl. Phys. {\bf  B263} (1986) 245; R.D. Ball,  Phys. Lett. {\bf B171}
(1986) 435; Phys. Rep. {\bf 182} (1989) 1; L. Alvarez-Gaume et al,
Phys. Lett. {\bf B166} (1986) 177 (and refs therein); A.Niemi and
G. Semenoff,  Phys. Rev. Lett. {\bf  55} (1985) 927.
\item{8.} R. Narayanan and H. Neuberger, Phys. Lett. {\bf B 302} (1993)
62; Nucl. Phys. {\bf B412} (1994) 574; Princeton prepr. IASSNS-HEP-94/99.
\item{9.} R. Narayanan and H. Neuberger, Phys. Rev. Lett. {\bf 71}
(1993) 3251; R. Narayanan, Nucl. Phys. B (Proc. Suppl.) {\bf 34} (1994) 95.

\end